\documentclass[prl,aps,a4paper,twocolumn,footinbib,superscriptaddress,showpacs,amsmath,amssymb]{revtex4}
\usepackage{graphicx}
\usepackage[
  colorlinks=true,
      linkcolor=blue,
      urlcolor=blue,
      filecolor=black,
      citecolor=blue,
            ]{hyperref}
\usepackage{color}
\usepackage{dcolumn}
\usepackage{bm}
\usepackage{slashed}
\def\d {\textmd{d}}

\begin{document}
\title{ Multi-horizon  and Critical Behavior  in Gravitational Collapse of Massless Scalar }

\author{Zhoujian Cao}
\email{zjcao@amt.ac.cn} \affiliation{Institute of Applied
Mathematics, Academy of Mathematics and Systems Science, Chinese
Academy of Sciences, Beijing 100190, China}
\author{Rong-Gen Cai}
\email{cairg@itp.ac.cn} \affiliation{Institute of Theoretical
Physics, Chinese Academy of Sciences, Beijing 100190, China}
\affiliation{School of Physics, University of Chinese Academy of
Sciences, YuQuan Road 19A, Beijing 100049, China}
\author{Run-Qiu Yang}
\email{aqiu@itp.ac.cn} \affiliation{Institute of Theoretical
Physics, Chinese Academy of Sciences, Beijing 100190, China}
\affiliation{School of Physics, University of Chinese Academy of
Sciences, YuQuan Road 19A, Beijing 100049, China}

\begin{abstract}

This paper studies the whole process of gravitational collapse and accretion of a massless scalar field in
asymptotically flat spacetime.  Two kinds of initial configurations are considered. One is the initial data without black hole,
the other contains a black hole. Under suitable initial conditions, we find that multi-horizon will appear, which means that
the initial black hole formed by gravitational collapse or existing at the beginning  will instantly expand and suddenly grow, rather than
grows gradually  in the accretion process.  A new type of critical behavior is found around the instant expansion. The numerical computation shows that the critical exponents
are universal.
\end{abstract}

\pacs{
  04.25.D-,   
  04.40.Dg,   
  98.62.Mw    
}

\maketitle


\paragraph*{Introduction.--}
In 1993, Choptuik~\cite{Choptuik} found a quite interesting  critical phenomenon in  gravitational collapse
of a massless scalar field in spherically symmetric, asymptotically flat spacetime with
a one-parameter family of initial data parameterized by $\epsilon$.  He found that there exists a critical value $\epsilon_*$ with a discrete
self-similar solution, when the initial data parameter $\epsilon$ is less than the critical one, the scalar field disperses into infinity and no black hole
forms; when $\epsilon$ is larger than the critical value, a black hole forms, its mass is proportional to the difference $\epsilon-\epsilon_*$
with a universal critical exponent.  Since then, similar critical behaviors have been found in various systems.
For a review, one can see Ref. ~\cite{Gundlach:2007gc}.
 In recent years, the gravitational collapse in the asymptotic anti-de Sitter (AdS) space-time has attracted a  lot of  attentions  as it shows a so-called ``weak turbulence''  effect~\cite{Bizon:2011gg} and gapped critical behavior~\cite{Olivan:2015fmy}.
All these critical phenomena are related to the
formation of black hole and are theoretically related to the
stability of the non-black-hole states~\cite{christodoulou1986problem}.

Once a black hole forms, it will accrete matters surrounding the black hole.
Accretion is  an ubiquitous phenomenon in our universe.  The accretion will
make black hole grow. Larger the accretion flow is, faster the
black hole grows. One interesting question is whether the black hole
always grows gradually or it is possible the accretion can make the
black hole expand instantly. In this paper, we will show that the
instant expansion of a black hole due to accretion may happen. Under
suitable conditions, the instant expansion behavior may happen
several times.
And more we find that the critical behavior appears
near the moment where the instant expansion just takes place. The
critical behavior can be expressed as a power law as the one found
by Choptuik~\cite{Choptuik}. Interestingly, the critical exponent is
universal in the sense that it is the same for different
configurations of initial data considered in this paper, the same for
different parameters involved in the physical process, the same for
different instant expansion events along a single evolution.

In particular, let us stress here that in the gravitational collapse of matter fields,
when an apparent horizon forms, some matters are still outside the apparent horizon. Some of these matters
will be accreted into the black hole, some will disperse into infinity if there is no reflecting wall outside the black hole.
However,  in the super-critical cases considered in Choptuik's study~\cite{Choptuik} and following works,
once an apparent horizon forms, the evolution is stopped due to the numerical calculation limit.
 In this sense the critical behavior found by Choptuik is just the  first
part of the whole story of gravitational collapse.   In this paper,
with our numerical code, we are able to study the whole process of
gravitational collapse. Although the final state is a stationary
black hole, the process is found rather rich unexpectedly. Multiple
horizons and trapped regions will form in the gravitational collapse
process. With the new horizons, we find a new critical behavior in
the gravitational collapse.  This critical behavior is the second
part of the story of gravitational collapse. And interestingly we
find that the critical exponents  of these two behaviors are the
same. Following \cite{Cai:2015pmv} we can define several mutual
related critical exponents for the new critical phenomenon. It is
shown that these critical exponents satisfy the universal scaling
laws proposed in \cite{Cai:2015pmv}.

\paragraph*{Physical Model and Numerical Method.--}

The physical model  we  are considering  is the
gravitational collapse of a real massless scalar field in
4-dimensional asymptotically flat space-time. The dynamics of the model is
governed by  Einstein-Klein-Gordon equations,
\begin{align}\label{Eins-scalar1}
  &G_{\mu\nu}=8\pi G[\partial_\mu\Psi\partial_\nu\Psi-\frac12g_{\mu\nu}(\partial\Psi)^2],\\
  &\nabla^2\Psi=0,
\end{align}
where $\Psi$ is the massless scalar field and $G_{\mu\nu}$ is the
Einstein tensor. In this paper, we will take the units with  $G=1$, and to simplify the numerical computation, we
impose the spherical symmetry.  The usual numerical
methods in Refs.
\cite{Choptuik,Bizon:2011gg,Cai:2015pmv,Garfinkle,Hamade:1995ce} and relavent papers
 are based on the simple spherical coordinates, which fail to
simulate the dynamics once an apparent horizon appears because of
the coordinate singularity, so that the later process of gravitational
collapse after the apparent horizon appears is still lack of well
investigation. In order to see the whole process of gravitation collpase, we will solve the
Einstein-Klein-Gordon equations~\eqref{Eins-scalar1} through the
Baumgarte-Shapiro-Shibata-Nakamura (BSSN) formulation
\cite{M.Shibata,Baumgarte}. The original BSSN approach is not
suitable for non-Cartesian coordinates. This issue can be remedied
by introducing a covariant version of the BSSN equations
\cite{Brown}.  This covariant BSSN formulation has been applied to
study the critical gravitational collapse in Ref.
\cite{Akbarian:2015oaa} and  to find the Choptuik's critical exponent and
self-similarity in the sub-critical case. Here we will first briefly review
the covariant BSSN formulation. For more details, one can refer to
Ref. \cite{Brown:2007nt}.  The  formulation  employs a 3+1
decomposition of the spacetime as,
\begin{equation}\label{metric1}
  ds^2=-\alpha^2\d t^2+\gamma_{rr}(\beta \d t+\d r)^2+r^2\gamma_{\theta\theta} \d\Omega^2,
\end{equation}
with 3-metric components $\gamma_{ij}$,  lapse function
$\alpha$ and shift vector $\beta$.
Differing from standard ADM formulation which rewrites Einstein's
equation as evolution equations for the 3-metric and the extrinsic
curvature $\{\gamma_{ij},K_{ij}\}$, the BSSN formulation decomposes
them  into conformal rescaling ADM dynamical variables,
\begin{equation}\label{BSSNdecomp}
  \gamma_{ij}=e^{4\phi}\tilde{\gamma}_{ij},~~K_{ij}=e^{4\phi}(\tilde{A}_{ij}+\tilde{\gamma}_{ij}K),
\end{equation}
where $e^{4\phi}$ is the conformal factor to make
$\det{\tilde{\gamma}_{ij}}=r^2\sin\theta$, $\tilde{A}_{ij}$ is the
conformally rescaled trace-free part of the extrinsic curvature and
$K=\gamma^{ij}K_{ij}$ is the trace of extrinsic curvature. Another
difference is the introduction of a conformal connection
$\tilde{\Lambda}^i=[\tilde{\Lambda},0,0]$, a true 3-vector, as an
primary dynamical variable in the covariant BSSN formulation. This
variable makes the resulted partial differential equations admit
strong hyperbolicity \cite{Hilditch:2013sba}. Thus in the
covariant BSSN formulation, the original dynamical variables in the ADM
formulation are transformed into $\{\phi, \tilde{\gamma}_{ij},
\tilde{A}_{ij}, K, \tilde{\Lambda}\}$.

In order to treat the physical singularity introduced by a black
hole, we use the moving puncture technique
\cite{Campanelli:2005dd,Baker:2005vv}. Together with
this numerical technique, the ``moving puncture'' coordinate
conditions are used. This gauge condition includes the ``1+log"
condition for the lapse function $\alpha$ and a version of a
Gamma-driver condition for $\beta$ \cite{C.Bona,Brown}.

In the covariant BSSN formulation in the spherical coordinates, Eqs.
\eqref{Eins-scalar1} are rewritten as 9 evolution equations for
gravity field with 2 first order evolution equations for matter
field $\Psi$ and its auxiliary variable $\xi\equiv\partial_t
\Psi-\beta \partial_r \Psi$. For sovling the total 11
evolutional equations, we need to specify suitable boundary
conditions at $r=0$ and $r\rightarrow\infty$, respectively. For the $r=0$
boundary, we distinguish the evolution variables into even ones and
odd ones. The even variables include $\phi$, $\tilde{\gamma}_{rr}$,
$\tilde{\gamma}_{\theta\theta}$, $\tilde{A}_{rr}$,
$\tilde{A}_{\theta\theta}$, $K$, $\alpha$, $\Psi$ and $\xi$, while
the odd variables are $\tilde{\Lambda}$ and $\beta$. Based on this
parity classification,  we use buffer point method to implement the
boundary condition \cite{Hilditch:2012fp}. As for the
$r\rightarrow\infty$ boundary we cut out the computation at some far
position and implement a Sommerfeld boundary condition there
\cite{Cao:2008wn}. We have confirmed through numerical test
that when the boundary position is far enough,  it will not
influence the results presented in this paper.

In order to set up the initial data we have to solve two constraint
equations (Hamiltonian and momentum). The matter field in this paper
is initialized by Gaussian type distribution and zero initial
velocity,
\begin{equation}\label{initialscalar}
  \Psi(0,r)=\epsilon\exp\left[-\frac{(r-r_0)^2}{\sigma^2}\right],~\partial_t\Psi(0,r)=0,
\end{equation}
where $\epsilon, r_0$ and $\sigma$ are parameters with $\epsilon,
\sigma>0$. The conformal metric and lapse function  are all
initialized to unity, i.e.,
$\tilde{\gamma}_{rr}(0,r)=\tilde{\gamma}_{\theta\theta}(0,r)=\alpha(0,r)=1$.
We impose  that the system has a time reversal symmetry, so that  the extrinsic
curvature, shift vector and conformal connection all vanish
$\tilde{A}_{rr}(0,r)=\tilde{A}_{\theta\theta}(0,r)=K(0,r)=\tilde{\Lambda}(0,r)=\beta(0,r)=0$.
Such a setting makes the momentum constraint be automatically satisfied.
With the initial setup, it is now only the value of conformal factor
unspecified. We obtain the  initial value by solving the Hamiltonian constraint. We adopt the puncture idea \cite{Brandt:1997tf} to
set
\begin{equation}\label{puncture-tans}
  e^{\phi(0,r)}=1+\frac{r_{h0}}{4r}+u(r),
\end{equation}
and solve the regular part $u (r)$. Here $r_{h0}$ is the horizon radius of an initial black
hole, which is taken to be zero if the initial configuration does not
contain  black hole.

\paragraph*{Black Hole Horizon and Instant Expansion.--}

To locate the black hole in the spacetime, we need find  horizon. 
We introduce two ``cross normalized" null vectors
$k_a=\frac{1}{\sqrt{2}}[\sqrt{\gamma_{rr}}\beta-\alpha,\sqrt{\gamma_{rr}},0,0]$
and
$\ell_a=-\frac{1}{\sqrt{2}}[\sqrt{\gamma_{rr}}\beta+\alpha,\sqrt{\gamma_{rr}},0,0]$.
Both null vectors are perpendicular to the symmetric sphere. And
$k_a$ is outward pointing and $\ell_a$ is inward pointing. Let
$\Theta_k$ and $\Theta_\ell$ be the expansions corresponding to $k_a$
and $\ell_a$, respectively. The apparent horizon (AH)  is the
2-spherical surface with $\Theta_k=0$. Note that when $\Theta_k=0$,
$\Theta_\ell=-\sqrt{2}a^{-1}\partial_r\ln(r\sqrt{\gamma_{\theta\theta}})$,
which is always negative. This shows that the 2-surface AH is  in fact
 always a marginally trapped surface and the collection of AHs
forms a trapped horizon (TH). Based on this fact, the surface with
$\Theta_k=0$ is TH and its foliation at a moment is an AH.

Now we present our numerical results. We first consider the
configuration  with $r_{h0}=0$, which means there is no  black hole
on the initial Cauchy surface.  Take an example, we fix $r_0=10,
\sigma=1$ and teat amplitude $\epsilon$ as the tuned parameter. For
small value of $\epsilon$, the energy will dissipate to infinity and
no black hole can form in the final state. There is a critical
amplitude $\epsilon_0$, when $\epsilon>\epsilon_0$, the black hole
can form and in the case of $\epsilon\rightarrow \epsilon_0^+$, the
initial apparent horizon radius (measured by the area $A$ of
symmetric sphere rather than the coordinate radius $r$)
$R_{AH}\equiv\sqrt{A/4\pi}$ and amplitude $\epsilon$ satisfy
the relation,
\begin{equation}\label{scaling1}
  R_{AH}\propto(\epsilon-\epsilon_0)^\beta
\end{equation}
with $\beta\approx0.38$. This is the critical behavior found by
Choptuik~\cite{Choptuik}. In most of relevant works, the numerical
simulation stops when the apparent horizon forms because of the
numerical instability introduced by the  singularity.
In this paper the covariant BSSN formulation enables us to give the
whole evolution of gravitational collapse. In the case with
$\epsilon\gtrsim \epsilon_0$, we find that the apparent horizon will
grow continuously and a final Schwarzschild black hole is left at
last.

\begin{figure}
\includegraphics[width=0.22\textwidth]{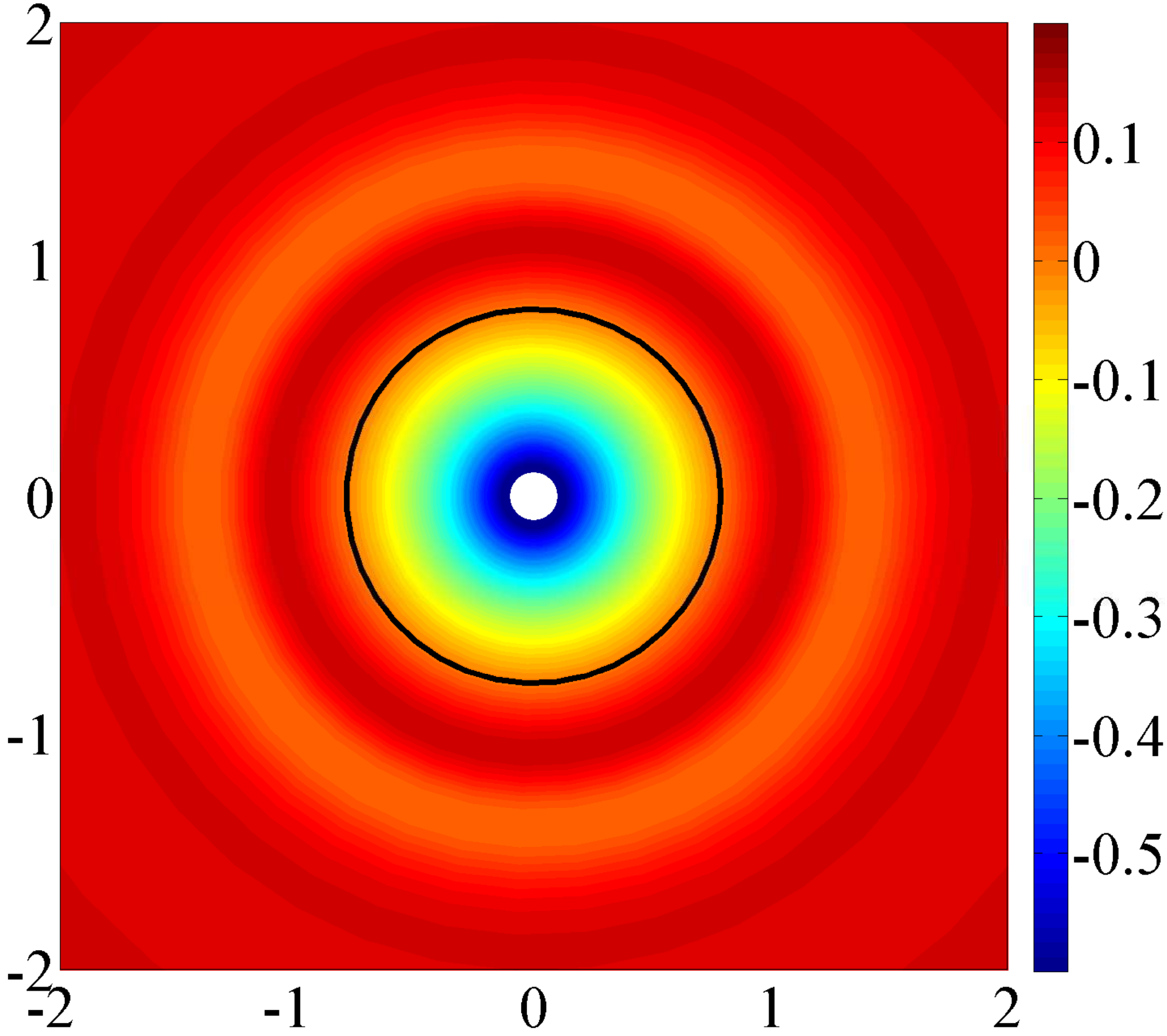}
\includegraphics[width=0.22\textwidth]{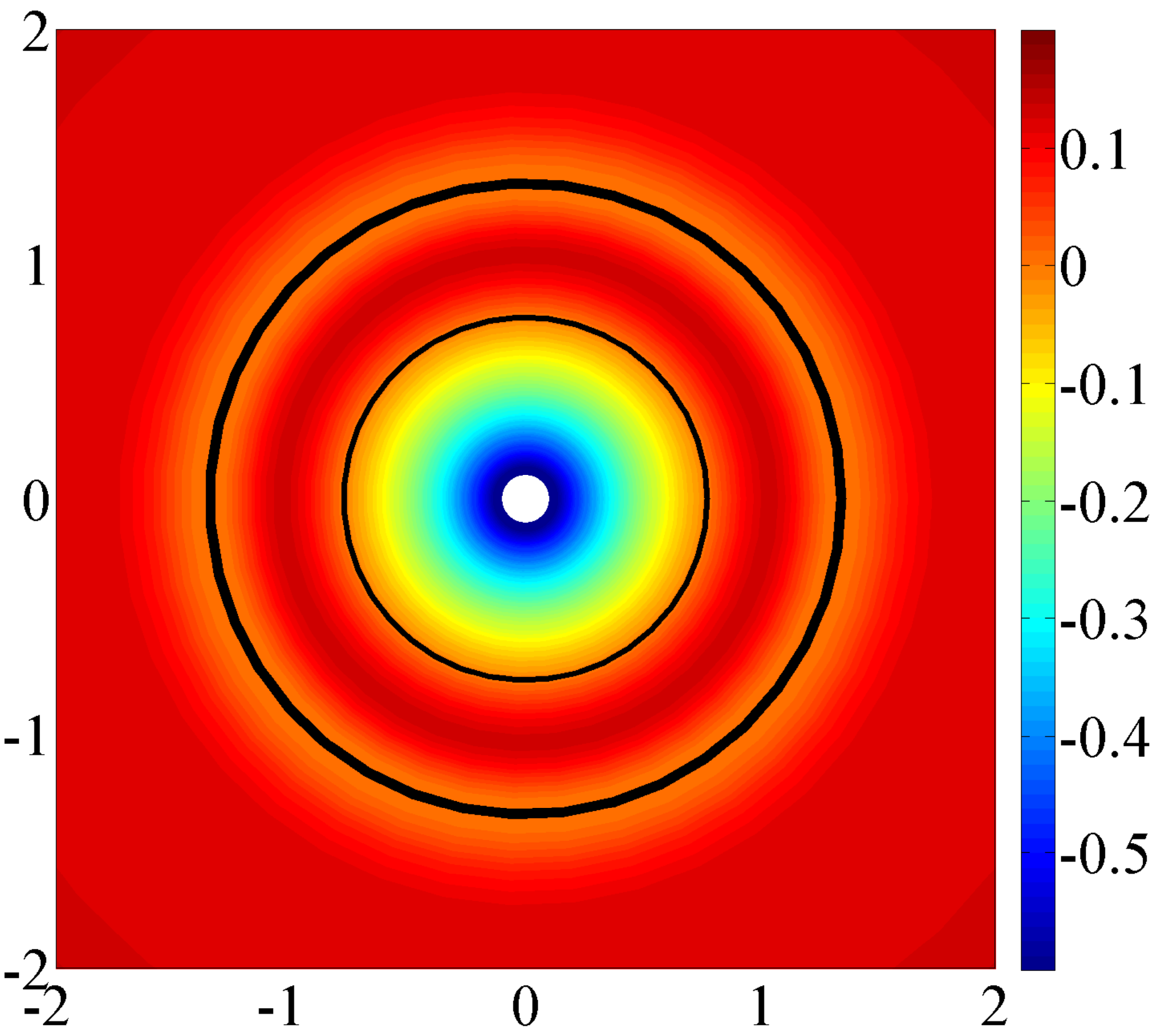}
\includegraphics[width=0.22\textwidth]{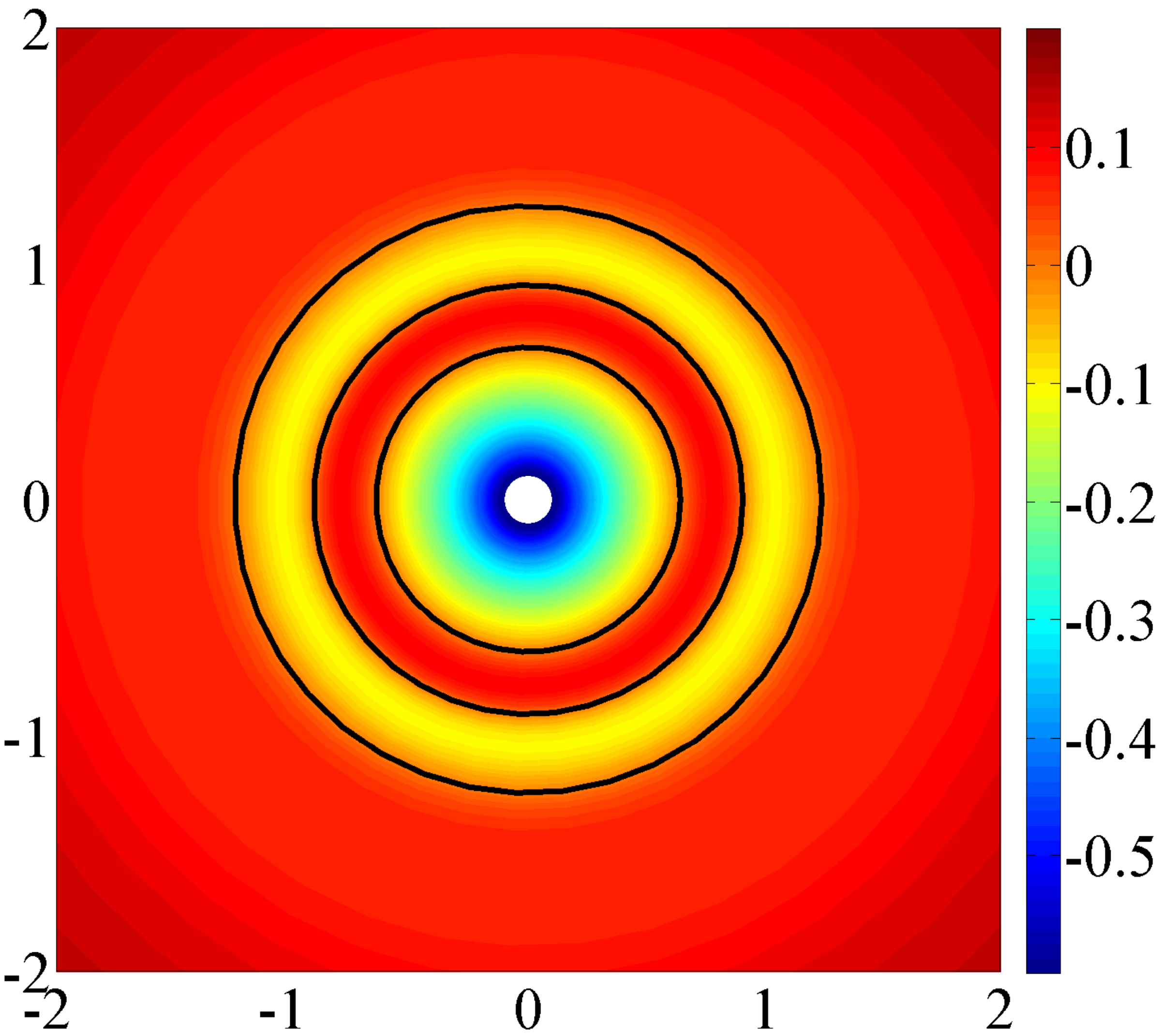}
\includegraphics[width=0.22\textwidth]{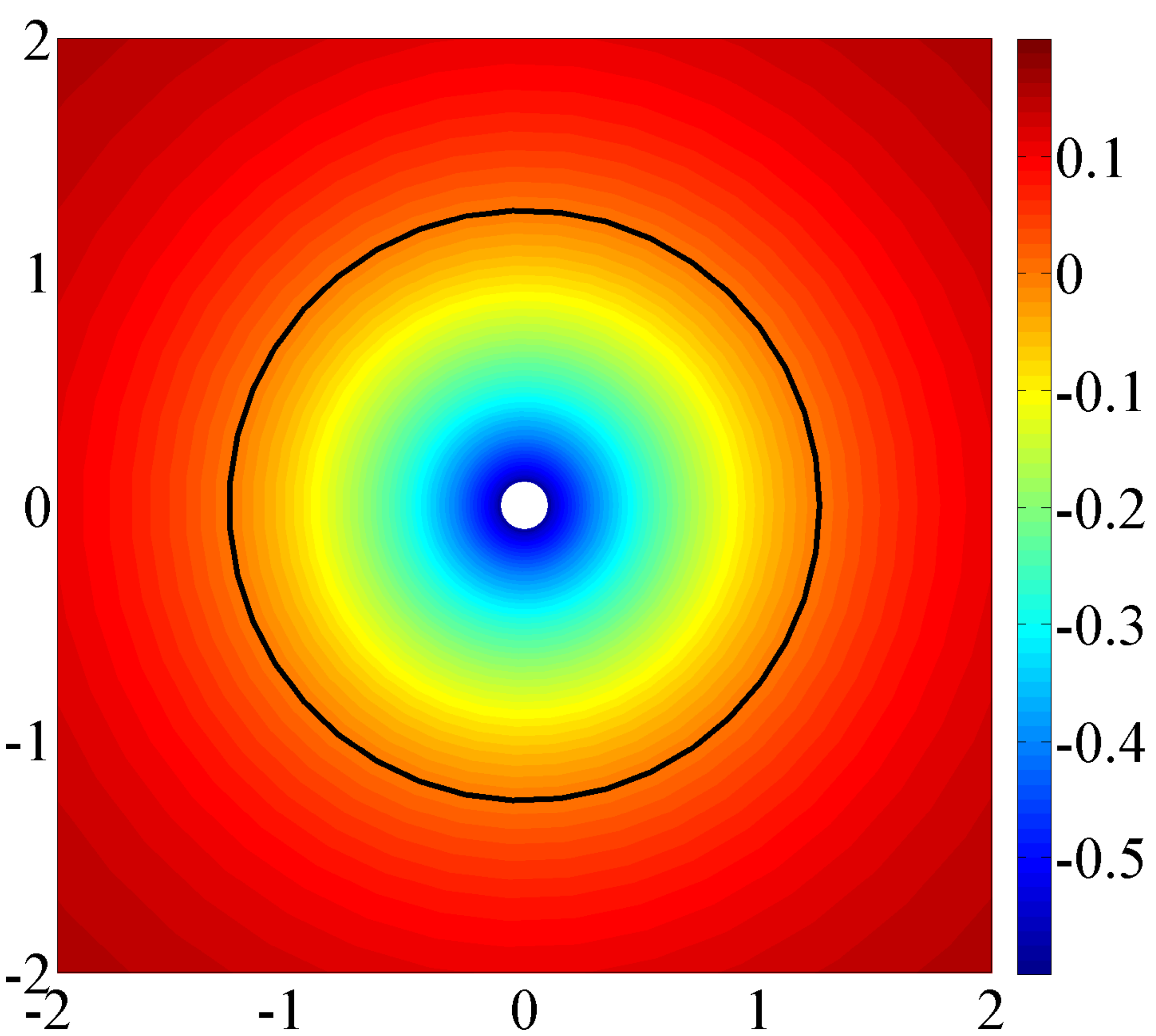}
\caption{The expansion $\Theta_k$ in equatorial plane for different
time. The black circles corresponds to the position $\Theta_k=0$.
The upper left panel corresponds to the time when the first AH
appears. The upper right panel corresponds to the time when the
instant expansion happens. The bottom left panel corresponds to the
time there are two trapped regions and three AHs. The bottom right
panel is the final state that only one AH left.} \label{expnasion4}
\end{figure}

Interesting thing happens when we increase the amplitude to strong
enough. We find that the black hole will expand instantly at some
time. This instant expansion is the result that  a new AH  appears
outside  the original AH. Since we treat the outmost AH as the
boundary of a black hole,  the black hole looks instant expansion,
namely the black hole  grows suddenly. One numerical result of
expansion $\Theta_k$ in the equatorial plane at different times is
shown in  Fig.~\ref{expnasion4}. It clearly shows that, after the
first AH appears, a new AH shows up outside the original AH  at some
time. Furthermore, this new AH splits into two AHs at  some time
late as shown in the bottom left panel of Fig.~\ref{expnasion4}.
There are two trapped regions (the yellow shell and the inner region
of the innermost circle) and a ``normal region" (the red shell)
between them. With further evolution, the ``normal region'" will
shrink and disappear,  which corresponds to the time that the two
inner AHs coincide with each other and they disappear. Finally only
the outermost AH is left.

\begin{figure}
\includegraphics[width=0.35\textwidth]{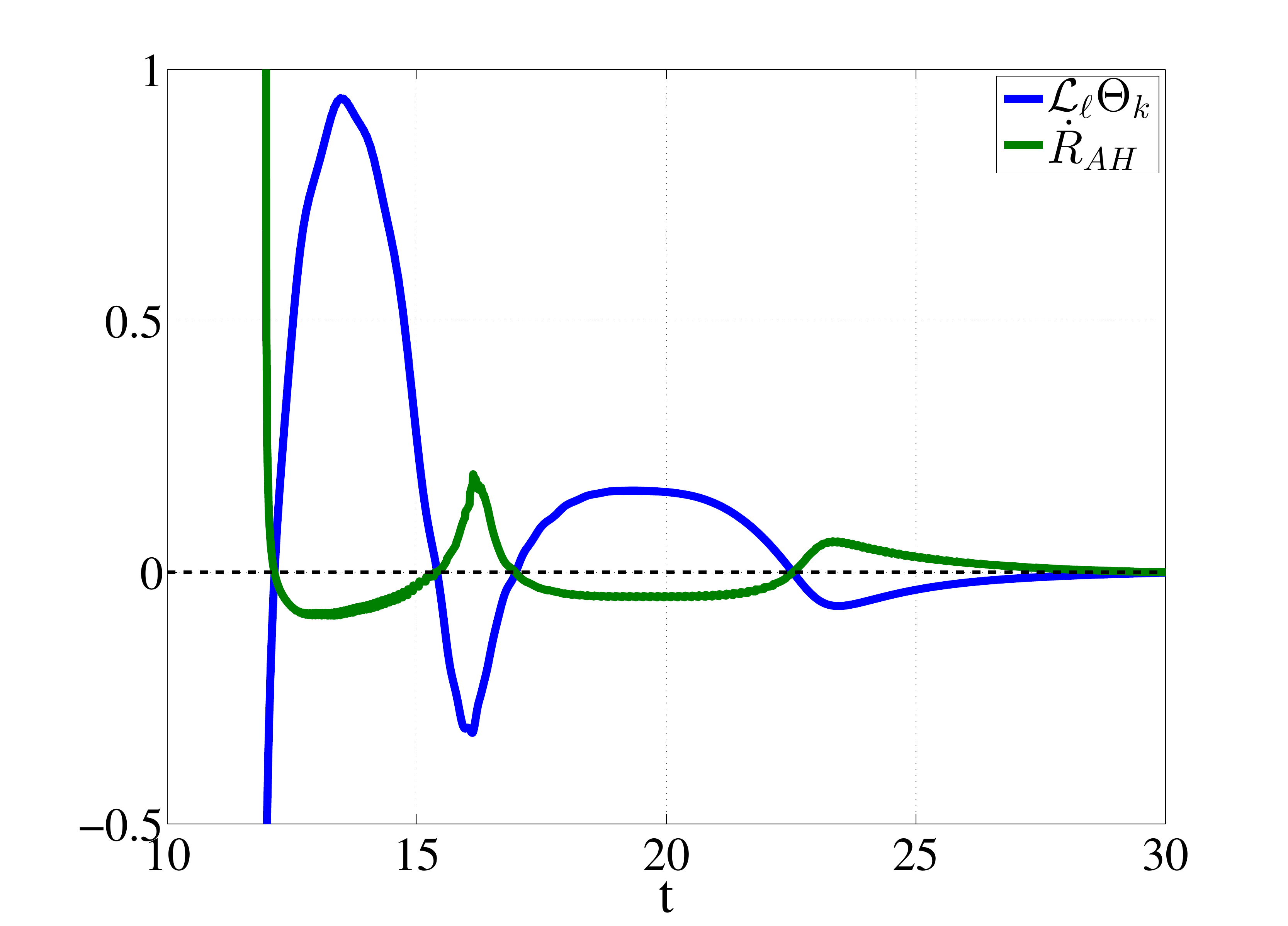}
\caption{The evolution of AH and the value of
$\mathcal{L}_{\ell}\Theta_k$ at the AH. Here $\dot{R}_{AH}$ is the
change rate  of the outmost apparent horizon radius. The parameters are taken as
$r_{h0}=0, \epsilon=0.04, r_0=10$ and $\sigma=1$. } \label{radiusAH}
\end{figure}

Unlike event horizon, the area of AH can increase or decrease during
the gravitational collapse. As shown in Fig.~\ref{radiusAH}, the
area of AH will increase when $\mathcal{L}_{\ell}\Theta_k<0$,  while
decrease when $\mathcal{L}_{\ell}\Theta_k>0$. This can be understood
as follows. In dynamical spacetime, AH can be the foliation of
future outer trapped horizon (FOTH) for negative
$\mathcal{L}_{\ell}\Theta_k$  or future inner trapped horizon (FITH)
for positive $\mathcal{L}_{\ell}\Theta_k$. Since the massless scalar
field satisfies the null energy condition, the area will increase
for FOTHs but decease for FITHs in the spherically symmetric
spacetime \cite{Booth:2005qc,Ashtekar:2005ez}. We verify  this
property numerically as an example shown in  Fig.~\ref{radiusAH}.

Now let us consider the case with the initial Cauchy surface containing  a
black hole. As  expected, we find  that instant expansion of  black hole can also happen. At
first glimpse,  this conclusion can be  naively deduced from the results for the
case with no black hole in the initial configuration we just discussed above.
Because once a black hole forms in gravitational collapse, some matter is still
outside the black hole, the configuration can be viewed as the initial configuration with
a black hole.   But in detail there is a  difference between these two initial
configurations. For the configuration which does not contain a black
hole at the beginning, the size of the formed black hole depends
on the initial matter setting. Stronger the initial matter is,
bigger the formed black hole is and fewer matter is left outside  the
black hole. But for the configuration that the initial data contains
a black hole, the size of the black hole and the content of the
matter can be tuned freely. More explicitly, the configuration which
does not  contain a black hole admits three parameters
$\{\epsilon, r_0, \sigma\}$, while the configuration which contains a
black hole initially admits four parameters $\{r_{h0}, \epsilon,r_0,
\sigma\}$. Our numerical calculations show that instant expansion of
the black hole may happen when we adjust any one of the four
parameters. And the behavior is similar to the one shown in
Fig.~\ref{expnasion4}.

Many times of instant expansions may happen during a given evolution
process if the initial setting is suitablly chosen. For example, if we
continuously increase the amplitude $\epsilon$, then two times or
three times instant expansions and multiple apparent horizons can
appear. However, we cannot obtain infinite times instant expansion by
increasing the amplitude in general, because new horizons may appear
in the initial Cauchy surface. When we adjust other parameters, we
get similar results.

\paragraph*{Critical Behavior due to Accretion.--}

\begin{figure}
\includegraphics[width=0.35\textwidth]{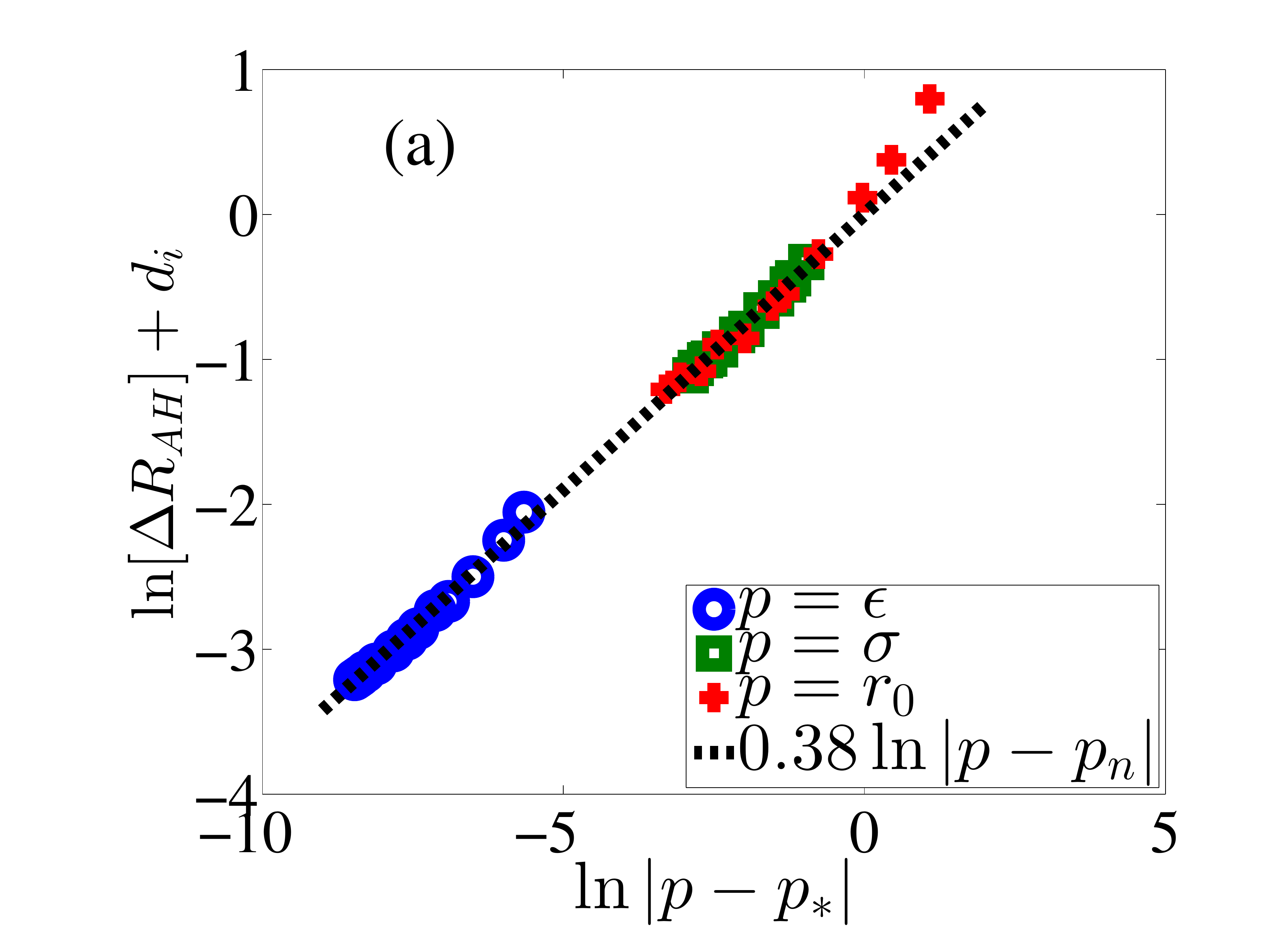}
\includegraphics[width=0.35\textwidth]{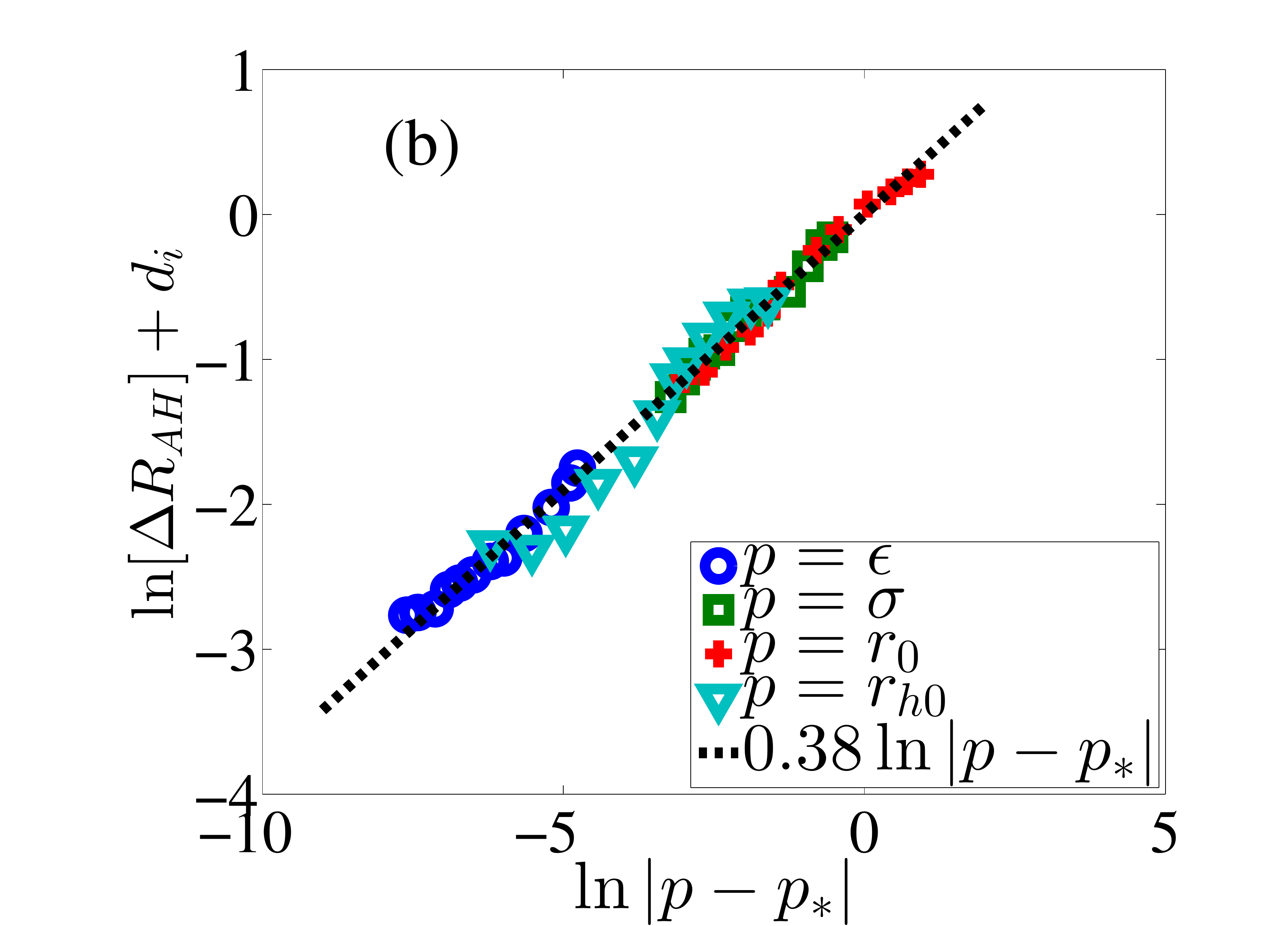}
\caption{The universal power law \eqref{scaling2} due to the
critical accretion for the first instant expansion during the given
evolution process. (a) For the initial configuration which does not contain
 black hole. (b) For the initial configuration which contains a
black hole. Here $d_i$ are imposed so that the intercepts
coincide with each other.} \label{powers1}
\end{figure}

After the first apparent horizon appears (formed from the
gravitational collapse or  imposed in the initial data), by
adjusting parameters involved in the initial configuration, a new
horizon can appear. When the instant expansion just happens (e.g.,
the moment shown in the up-right panel of Fig.~\ref{expnasion4}), we
find that the area radius difference between the two apparent
horizons $\Delta R_{AH}$ obeys a universal scaling law. Let $p$
stand for any tuning parameter in the initial configuration and
$p_*$ denote the critical value that a new horizon can just appear
when we fix the other parameters, the power law can be expressed as,
\begin{equation}\label{scaling2}
  \Delta R_{AH}=a_p|p-p_*|^{\beta_p}, ~~\text{when}~p\rightarrow p_*^+.
\end{equation}
Here $a_p$ is some constant and $\beta_p$ is the critical exponent which is
independent of the choice of $p$, independent of the initial
configurations and independent of which instant expansion during the
given evolution process. We have computed $a_p$ and $\beta_p$ for
different tuning parameters in both two configurations.  For
the initial configuration which dose not contain a black hole, we find  a
set of critical values $\{r_{0*}, \epsilon_*, \sigma_*\}$. After
tuning any one of them but fixing  others, we can obtain the results of
$\Delta R_{AH}$. The results are shown in Fig.~\ref{powers1}(a).
On the other hand, for the initial configuration which contains a  black
hole, we also find a set of critical values $\{r_{h0*}, r_{0*},
\epsilon_*, \sigma_*\}$. A similar relation between $\Delta
R_{AH}$ and tuned parameters is shown in
Fig.~\ref{powers1}(b). For both configurations, we find an universal
exponent $\beta_p\approx0.38$. Besides the first instant expansion,
for the case that the two times or three times instant expansion can
happen during the accretion, we also find a similar power law
behavior for the outmost horizon and its nearest horizon when the
outmost horizon just appears. The critical exponents are the same up
to the numerical error.

As a self-consistency  check, we compute the `susceptibility' between
any two tuning parameters $p$ and $p'$ as $\chi_{p'p}=(\partial
p'_*/\partial p)_{\Delta R_{AH}=0}=c_{p'p}|p-p_*|^{\delta_{p'p}}$.
We find $\delta_{p'p}\approx0$ and exponents and coefficients
in Eq.~\eqref{scaling2} satisfy following two equations proposed in
Ref.~\cite{Cai:2015pmv},
\begin{equation}\label{relations1}
  \beta_p(\delta_{pp'}+1)\approx\beta_{p'},~~~a_{p'}|c_{p'p}|^{\beta_p}/a_{p}\approx1.
\end{equation}

Although our new scaling behavior Eq. \eqref{scaling2} looks similar to the one
found by Choptuik, there is an essential difference between them.
In the  Choptuik's case, the scalar curvature diverges at the critical point.
However, the new critical behavior found here happens in a regular region, where all
the geometrical quantities are finite. In other words, the Choptuik's scaling law is applicable
for a small black hole case, while ours is also valid for a finite mass black hole.

\paragraph*{Discussion.--}

 Our computations in this paper are  done in the asymptotically flat spacetime,
it can be  straightforwardly generalized to the asymptotic AdS case in
principle.  Due to the ``weak turbulence" effect and the existence of the reflection
boundary of AdS space, we expect that  rich structures in
the inner region of black hole and criticality between multiple
horizons can appear in the gravitational collapse in AdS spacetime.

Note that the Choptuik's critical exponent can be obtained by analyzing the
Lyapunov's exponents and renormalization group of the system~\cite{Koike:1995jm},
it is interesting to consider if the scaling behaviors in
Eq.~\eqref{scaling2} can also be obtained in a similar way. As we have
mentioned, since  there is an essential difference between this case and
Choptuik's case, such a work will not be a trivial generalization.  Numerically, we find that critical exponents
$\beta_n$ here is roughly the same as the Choptuik's critical
exponent~\cite{Akbarian:2015oaa}. Do they really take the same
value? If yes, why do such different critical behaviors have the
same critical exponent? We hope all these interesting questions could
be addressed in  future.

To conclude, we have numerically investigated the accretion process of scalar field
by a black hole which is formed through gravitational collapse
or exists at the beginning.  With the covariant BSSN formulation, we are able
to study the whole process of the gravitational collapse.
For both configurations, we have found that the
black hole will undergo instant expansion during the accretion under
very wide conditions. The appearance of the new apparent horizons is
due to the weight enough of the mater outside the original horizon to form
new horizon.  A new universal power law behavior was found
in the instant expansions. The critical exponent is the same not
only for different one-parameter initial configurations but also for
all the instant expansions. Since accretion phenomenon is quite
ubiquitous in the universe, it would be quite interesting to see whether the existence of the intermediate  and supermassive black holes is
 related to the critical behavior found in this paper.

\paragraph*{Acknowledgments.--}
This work was supported in part by the National Natural Science
Foundation of China (No.~11375247, No.~11375260 and No.11435006).



\end{document}